**On the origin of red luminescence from iron-doped $\beta$-$Ga_2O_3$ bulk crystals**

Rujun Sun[1*], Yu Kee Ooi[1*], Peter T. Dickens[3], Kelvin G. Lynn[3], Michael A. Scarpulla [1,2]

[1.] *Electrical and Computer Engineering, University of Utah, Salt Lake City, UT, 84112, USA*

[2.] *Materials Science and Engineering, University of Utah, Salt Lake City, UT, 84112, USA*

[3.] *Materials Science & Engineering Program, Washington State University, Pullman, WA, 99164, USA*

**Abstract**

Currently, Fe doping in the ~$10^{18}$ $cm^{-3}$ range is the most widely-available method for producing semi-insulating single crystalline $\beta$-$Ga_2O_3$ substrates. Red luminescence features have been reported from multiple types of $Ga_2O_3$ samples including Fe-doped $\beta$-$Ga_2O_3$, and attributed to Fe or $N_O$. Herein, however, we demonstrate that the high-intensity red luminescence from Fe-doped $\beta$-$Ga_2O_3$ commercial substrates consisting of two sharp peaks at 689 nm and 697 nm superimposed on a broader peak centered at 710 nm originates from Cr impurities present at a concentration near 2 ppm. The red emission exhibits two-fold symmetry, peaks in intensity for excitation near absorption edge, seems to compete with $Ga_2O_3$ emission at higher excitation energy and appears to be intensified in the presence of Fe. Based on polarized absorption, luminescence observations and Tanabe-Sugano diagram analysis, we propose a resonant energy transfer of photogenerated carriers in $\beta$-$Ga_2O_3$ matrix to octahedrally-coordinated $Cr^{3+}$ to give red luminescence, possibly also sensitized by $Fe^{3+}$.

* both authors contribute equally to this paper

*Corresponding Author: Michael A. Scarpulla, email: mike.scarpulla@utah.edu*

$\beta$-$Ga_2O_3$ exhibits an ultrawide bandgap from 4.5 to 4.8 eV (with optical axis dependent) suggesting its application for electronic devices requiring high breakdown field[1]. Both widely-variable n-type conducting and semi-insulating layers can be achieved with extrinsic doping. Specifically, controllable n-type doping can be achieved in multiple growth techniques using Si dopant as well as Ge, Sn, Zr and Hf. Dopants of Fe, Mg and N as deep acceptors are used to construct $\beta$-$Ga_2O_3$ FET lateral devices, current blocking layers in vertical devices, and to provide highly-resistive substrates/buffer layer for other applications[2,3]. Computations show that oxygen vacancy is deep donor, while gallium vacancies and their various complexes, especially with hydrogen, are the most probable dominant compensating native defects[4-8].

Fe-doped $\beta$-$Ga_2O_3$ semi-insulating crystals are commercially available and widely used as substrates for epitaxial $\beta$-$Ga_2O_3$ layers. In these crystals, the Fe dopant concentration is around $10^{18}$ $cm^{-3}$ to compensate background donors and pin the Fermi level[9-11]. The band gap of Fe-doped $\beta$-$Ga_2O_3$ has reported as ~4.5 eV[12], and reduced from ~4.6 eV to ~2.9 eV with [Fe]/([Ga]+[Fe]) from 0.0 to 0.4[13]. The deep donor-like level (E2) of Fe is measured as $E_c$-0.78 eV using deep level transient spectroscopy[14]. The $Fe^{2+/3+}$ charge transition level has determined as $E_c$-0.84$\pm$0.05 eV using noncontact spectroscopy methods (DLTS)[15]. However, other work has reported the optically-induced change from $Fe^{3+}$ as 1.3±0.2 eV using steady-state photo-induced electron paramagnetic resonance (EPR) measurements[16]. For luminescence, Polyakov et al. reported two sharp emission lines near 1.78 eV and 1.80 eV at low temperature[17] suggesting an origin from highly-localized atomic states (e.g., d orbitals). They ascribed the 1.78 eV peak to the $^4T_1 \rightarrow {}^6A_1$ intracenter transition of $Fe^{3+}$. This assignment was based on a logical but circumstantial argument that since Fe is the highest-concentration intentional impurity, this emission is probably related to Fe. Hany et al. reported that a red to near-infrared band (R-NIR) emerged after annealing in the air, and two extremely sharp $R_1$ and $R_2$ peaks appeared below 140 K[12]. The R-NIR sharp peaks were ascribed to nitrogen incorporated during air annealing. However, this is also circumstantial and at odds with the general finding that transitions arising from states mixed with β-$Ga_2O_3$ matrix states are broadened by the strong carrier-lattice coupling[18]. Lastly, similar red peaks are observed in thermally stimulated luminescence of UID[19], Mg-doped[20] and Fe-doped[21] $\beta$-$Ga_2O_3$ crystals and electroluminescence of Si and Cr co-doped $\beta$-$Ga_2O_3$ crystals[22] and origins from Fe and Cr are claimed but also lacking detailed investigation.

Here we seek to clarify some details of the red and near-infrared luminescence from Fe-doped $\beta$-$Ga_2O_3$ crystals. Reduced optical band gap and increased sub-bandgap absorption are observed. In PL, $\beta$-$Ga_2O_3$ emissions are strongly quenched and an additional structure emerges of a broad red peak around 710 nm with two sharp peaks at 688 nm and 696 nm. The sharp and broad red peaks are assigned to emission of $^2E \rightarrow {}^4A_2$ and $^4T_2 \rightarrow {}^4A_2$ of Cr internal transitions, respectively. Finally, we discuss the possible origins of the Cr co-doping and possible mechanisms within the luminescence pathway.

Three types of Fe-doped crystals [Syn (100) and Syn (010) grown by the Czochralski (CZ) method by Synoptics, and NCT (010) grown by the edge-fed growth method by Novel Crystal Technology were studied. Additionally, we measured an (100) unintentionally-doped (UID) crystal grown at Washington State University (WSU) by vertical gradient freeze (VGF). We use polarization-dependent transmission measurements to determine bandgaps and absorption coefficient. Photoluminescence was collected using a fiber coupled spectrometer in an integrating sphere as a function of polarization of the incident laser. For temperature-dependent PL, the laser beam was normal to the sample surface while the luminescence was collected with an optical fiber located perpendicular to the laser beam. We quantified the concentrations of Fe and Cr using ICPMS.

The fundamental absorption of $\beta$-$Ga_2O_3$ occurs from the valence band maxima composed mostly of O-2p orbitals to the conduction band minima composed of Ga-4s orbitals[23]. The absorption edges for different axes result from selection rules and splitting related to the O-2p states. Fig. 1a and 1b show the transmittance vs. polarization angle of Syn (100) and Syn (010) Fe-doped $\beta$-$Ga_2O_3$, respectively. For Syn (100) sample, the absorption onset rises sharply with photon energy. The optical transition thresholds deduced from Tauc plots of $(\alpha h\nu)^2$ vs. $h\nu$ for E||c and E||b are 4.54 eV and 4.81 eV, respectively, in good agreement with prior reported values[23]. For the Syn (010) sample, the absorption slowly increases with photon energy for each incident direction, and the band gap of E||c and E||a* are 4.25 eV and 4.34 eV, respectively. The ordering and energies of these thresholds are also consistent with literature: $E_{g,\,E||c} < E_{g,E||a*} < E_{g,E||b}$ [24]. The reduction of the apparent threshold for E||c and E||a* of the Syn (010) sample could be explained by larger concentration of Fe and associated disorder[13]. A peculiar feature of the transmittance data is that the data for all incident angles cross at one particular energy near 4.30 to 4.40 eV for both samples.

This crossing of the data may result from changes in reflectivity near the band edge associated with the birefringence and changes in the refractive index for the two directions near the optical absorption transition.

Moreover, in Fig. 1c, we observe anisotropy in the sub-bandgap absorption. Both samples' data illustrate regions that show exponential energy deference, i.e., $\alpha = \alpha_0 \exp\left(\frac{E-E_g}{E_U}\right)$, where $\alpha_0$ is a constant, $E$, $E_g$, $E_U$ are photon energy, optical band gap and Urbach energy, respectively. Fitting using the above expression yields $E_U$ = 76 meV and 52 meV for $E$||b and $E$||c, respectively, for the Syn (100) Fe-doped sample. These are considerably larger values as compared for example to crystalline Si and GaAs (11-18 meV), but are in the range for hydrogenated amorphous Si (50-100 meV)[25]. Furthermore, $E_U$ for the Syn (010) Fe-doped sample are 600 meV and 360 meV for $E$||a* and $E$||c, respectively (Fig. 1c). These values are similar to the levels found for ion-implanted GaAs (300-520 meV)[25]. The typical physics of positional disorder induced Urbach energies would be expected to produce isotropic sub-gap absorption, whereas the fact that we observe anisotropy seems to suggest the transitions are still tied to specific anisotropic or selection rule features in the band structure or anisotropic defect absorption. Therefore, we consider it unlikely that the true Urbach parameter is so large given that the samples are crystalline. It is possible that the slopes of the data in this region reflect complications of the shape of the absorption edge by the dipole-forbidden minimum bandgap transition which is close in energy[26]. Lastly, we note that no additional absorption bands are detectable across the wider visible range for both the Syn (010) and Syn (100) samples (Fig. 1d). This shows that these samples contain low enough concentrations of transition metals that their intracenter absorption bands are below detection limits.

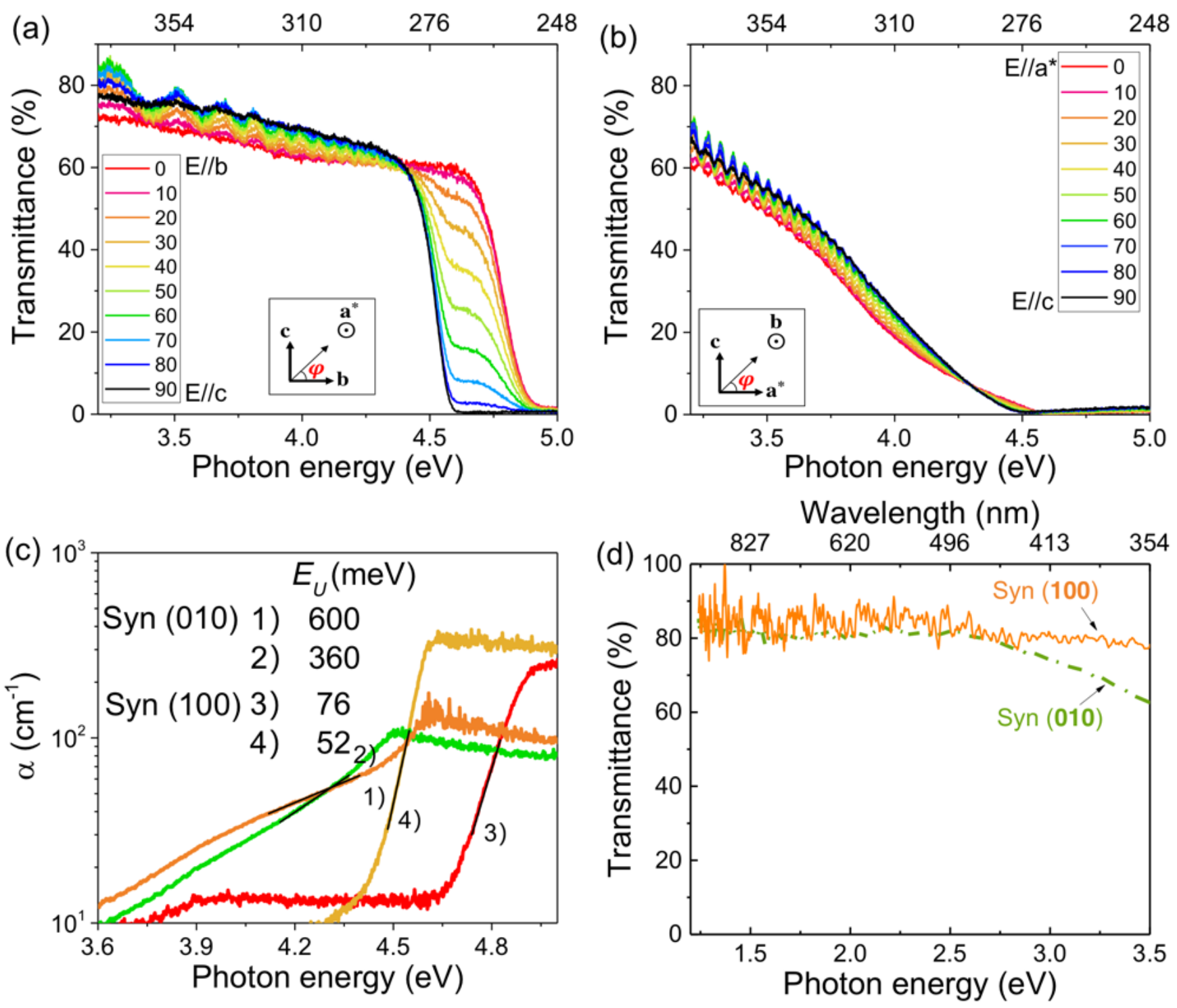

Figure 1 - Transmittance spectra for different light polarization angles for (a) Syn (100) and (b) Syn (010)-oriented Fe-doped $\beta$-$Ga_2O_3$ crystals. (c) Estimated absorption coefficient corrected by refractive index and its Urbach tail fitting. (d) Un-polarized transmittance over a wider sub-gap energy range for Syn (100) and Syn (010) samples.

Fig. 2a show typical PL spectra for NCT (010), Syn (100), and Syn (010) Fe-doped $\beta$-$Ga_2O_3$. The UID sample shows the typical UV, UV', blue, and green emissions[27]. The Fe-doped samples show UV', blue, and green peaks of ~100 times lower in intensity as compared to those from the UID samples. Additionally, the dominant red emission from these Fe doped samples consists of a broad peak centered at 710 nm with two sharp peaks at 689 nm and 697 nm ($R_1$ and $R_2$) superimposed. Low-temperature PL shows that the two sharp peaks intensify while the broad peak decreases and eventually diminishes as the temperature decreases (Fig.2b). The quenching of the broad peak at low temperatures indicates that its emission requires an excitation over a thermal barrier. The two sharp peaks exhibit very narrow bandwidth (FWHM <3 nm for all measured temperatures). The energy difference between the two sharp peaks is 18.1 meV at around 100 K. Red emission shows 2-fold symmetry which coincides with incident polarization parallel to a*, b and c axes. This implies that the optical absorption mainly occurs in $\beta$-$Ga_2O_3$ matrix. The photoluminescence excitation (PLE) spectra for both the UID WSU (100) and the Syn (100) Fe-doped sample show similar trends in E//b direction (Fig. 2c and 2d) but differ in E//c. The intensity of UV emission from UID sample drops continuously with increasing excitation energy ($E_{ex}$). The intensity of 689 nm emission of Syn (100) sample fades above 4.9 eV, then the emission intensity integrated from 300 to 600 nm from $\beta$-$Ga_2O_3$ emerges. The red emission in E//c is inefficient above 4.9 eV excitation and $\beta$-$Ga_2O_3$ emission occurs simultaneously, suggesting a competition for photocarriers between these two emission pathways.

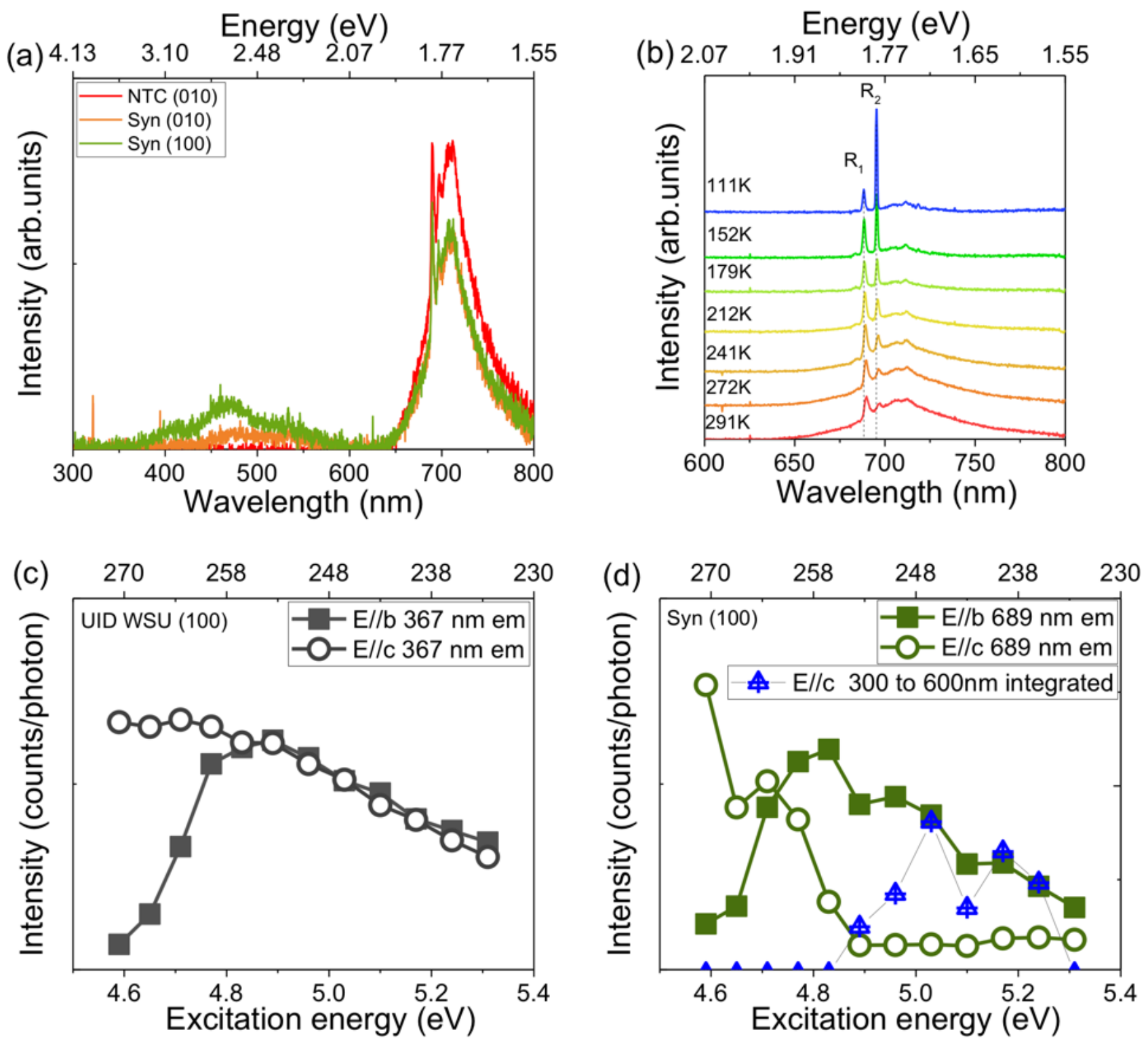

Figure 2 – (a) PL spectra in arbitrary scale for Fe-doped crystals under excitation of 266 nm. (b) representative temperature-dependent PL spectra for Fe-doped $\beta$-$Ga_2O_3$ crystal. (c) PLE for UID WSU (100) crystal observed at 367 nm; (d) PLE spectra for Syn (100) Fe-doped $\beta$-$Ga_2O_3$ observed at 689 nm, blue triangle represents integrated intensity from 300 nm to 600 nm at E//c.

The red PL emission originating from nitrogen incorporation[28] differs in three aspects as compared to these from Fe-doped $\beta$-$Ga_2O_3$ bulk crystal: 1) the centroid shifts from 1.71 eV to 1.65 eV with temperature decreasing to 10 K; 2) nitrogen related peak exhibits only one broad peak from 300 to 10 K (with FWHM ~ 0.4 eV); 3) this broad peak intensifies with decreasing temperature rather than disappearing.

Hybrid functional calculations show that self-trapped holes and many extrinsic defects can give rise to very broad luminescence bands[18] due to strong electron phonon coupling. This applies for acceptors such as Mg, Ca and N, shallow donors of Si and Sn, and other impurities like B, Na, Al, S, Cl, Pb and Bi if they are involved in PL. We stress here that is the expectation for emission arising from states that mix significantly with the states of the $\beta$-$Ga_2O_3$ matrix, as opposed to those arising from atomic-like d and f shell transitions. The narrow bandwidths of the $R_1$ and $R_2$ peaks in Fe-doped $\beta$-$Ga_2O_3$ suggest that it arises from internal transitions of ions, specifically, spin-forbidden transitions with long lifetime. The spin-allowed transitions exhibit fast decay leading to lifetime-broadened peak peaks[29]. we are well aware that many extrinsic impurities including some transition metals are found in $\beta$-$Ga_2O_3$ melt-grown crystals, some originating in the feedstock materials and some introduced during growth (especially from crucibles). Transition metal impurities in $\beta$-$Ga_2O_3$ crystals include Ir, Zr, Ti and Ni [30]. These elements are not believed to contribute to the sharp red emission structure we discuss here in this paper. Transitions related to all mentioned but the Fe and Cr are summarized in **Table 1**. Ir has not been observed as a luminescent center in minerals and semiconducting or insulating compounds, although some of its complexes do luminesce.

**Table 1** Ir, Zr, Ti and Ni related emission properties in octahedral Ga sites

| Elements | Electron configuration | Transitions | Note |
|---|---|---|---|
| $Ir^{4+}$ | $[Xe]4f^{14}5d^{5}$ | $^2T_2$ (I) split by spin-orbital coupling: expected sharp peak at 0.64 eV[31,32] | Spin-forbidden |
| $Zr^{4+}$ | [Kr] | – | – |
| $Ti^{3+}$ | $[Ar]3d^{1}$ | $^2E \rightarrow {}^2T_2$: broad peak at 1.64 eV[33,34] | Spin-allowed |
| $Ni^{2+}$ | $[Ar]3d^{8}$ | $^3T_2 \rightarrow {}^3A_2$: broad peaked at 0.86 eV[35] | Spin-allowed |

Now we discuss internal transitions from $Cr^{3+}$ and $Fe^{3+}$ in both β-$Ga_2O_3$ and $Al_2O_3$ (corundum/ sapphire). Tanabe-Sugano diagrams predict the dependence of transitions energy on the perturbing crystal field imposed on transition metals by the nearest-neighbor atoms treated as point charges. This allows determination of the crystal field strength and predicting unobserved transitions using known absorption bands/lines. Both experiment and theory calculations indicate that Fe and Cr in $\beta$-$Ga_2O_3$ show a preference for incorporation on the octahedrally-coordinated $Ga_{II}$ site[14,36]. The Tanabe-Sugano diagrams for $d^3$ and $d^5$ electron configurations of $Cr^{3+}$ and $Fe^{3+}$ in octahedral coordination are presented in Fig.3[37].

Two absorption bands at 428 nm and 600 nm in Cr-doped $\beta$-$Ga_2O_3$ crystals (> 0.1 at%)[38,39], correspond to the $^4A_2(F)\rightarrow{}^4T_1(F)$ and $^4A_2(F)\rightarrow{}^4T_2(F)$ absorption transitions, which implies the $\frac{\Delta_0}{B}$ parameter is near 24.8. This value is close to 23.1-23.5[40,41] reported for other Cr-doped $Ga_2O_3$ crystals. Interestingly, this value of B predicts that another absorption transition for $^4A_2(F)\rightarrow{}^4T_1(P)$ should exist near 271 nm (4.58 eV), which is nearly resonant with the fundamental absorption of $\beta$-$Ga_2O_3$ for E//a* and c. This value of $\frac{\Delta_0}{B}$ determined from the absorption implies that the $^2E(G)\rightarrow{}^4A_2(F)$ emission band should be near 693 nm, which is very close to the sharp red emission we observe (at 689 and 696 nm in Fig. 2). For comparison, Cr in $Al_2O_3$ exhibits absorption bands at 400 nm and 555 nm[42,43] indicating $\frac{\Delta_0}{B}$ of 28.0 because of slightly smaller bond lengths. The well-characterized ruby $R_1$ and $R_2$ emission lines at 692 nm and 694 nm in $Al_2O_3$ result from the splitting of $^2E$ due to a combination of the crystal field and spin-orbit interaction[44]. The Tanabe-Sugano diagrams in Fig. 3 show that the $^4A_2\rightarrow{}^4T_1$ and $^4A_2\rightarrow{}^4T_2$ transition energies change significantly with crystal field strength, but that of the $^2E\rightarrow{}^4A_2$ transition does not. Thus, the experimental observation that the absorption bands in Cr doped $Al_2O_3$ and $\beta$-$Ga_2O_3$ differ while the emission transitions are very close in energy, is consistent with the assignments above. The broad red luminescence comes from $^4T_2\rightarrow{}^4A_2$ due to a partial thermal population from $^2E$ to $^4T_2$ [45]. With decreasing temperature, the thermal population of $^4T_2$ states is suppressed, and thus the broad peak disappears as we observe.

We now examine the possibility that the red luminescence might arise from Fe in $\beta$-$Ga_2O_3$. $Fe^{3+}$ absorption lines are confirmed by the yellowish color of crystals and distinct absorption around 455 nm for $^6A_1(S)\rightarrow{}^4A_1(G)/{}^4E(G)$ which is rather independent of crystal field for high-spin $Fe^{3+}$ regime[13,46]. The $\frac{\Delta_0}{B}$ of octahedral Fe in $\beta$-$Ga_2O_3$ single crystals is calculated as 21 using the two absorption bands at 460 nm and 689 nm[13] corresponding to $^6A_1(S)\rightarrow{}^4A_1(G)$ and $^6A_1(S)\rightarrow{}^4T_2(G)$ transitions, respectively. This restricts its emission from $^4T_1\rightarrow{}^6A_1$ to be at 980 nm (1.265 eV), which is very far away from the red emissions near 700 nm (1.75 eV) that we (and others) observed from Fe-doped $\beta$-$Ga_2O_3$. Similarly, this value of B predicts that another possible absorption transition for $^6A_1\rightarrow{}^4A_2(F)$ near 278 nm (4.46 eV). This is nearly resonant with $\beta$-$Ga_2O_3$ matrix and with the Cr absorption discussed early. The luminescence of $Fe^{3+}$ doped $\alpha$-$Ga_2O_3$ powder is observed at 950 nm (1.305 eV)[47] where $\alpha$-$Ga_2O_3$ , just like corundum/ruby $\alpha$-$Al_2O_3$, contains only octahedrally-coordinated cation sites. Lastly, $\frac{\Delta_0}{B}$ for octahedral $Fe^{3+}$ in $Al_2O_3$ is 22.0-22.9[46,48] predicting emission band for the $^4T_1\rightarrow{}^6A_1$ transition be close to 1050-1130 nm (1.181-1.097 eV). Therefore, we can conclude that red luminescence originating from $Fe^{3+}$ in Fe-doped $\beta$-$Ga_2O_3$ is highly unlikely.

Note that $Fe^{3+}$, $Cr^{3+}$ and $Ir^{3+}$ can form $Fe^{2+}$, $Cr^{4+}$ and $Ir^{4+}$, respectively, by capturing photogenerated electrons and holes. Since the Fermi energy is fixed at $Fe^{3+}/Fe^{2+}$ level (~ $E_c$–0.8 eV) for Fe-doped $\beta$-$Ga_2O_3$ semi-insulating samples (for n-type doped sample, $E_F$ is near $E_c$ due to shallow doping), most of Cr and Ir (deep donors) are in neutral states, namely, $Cr^{3+}$ and $Ir^{3+}$. Thus only a small part of these ions change charges[15]. Besides, none of them has observed to give sharp red emission near 1.8 eV.

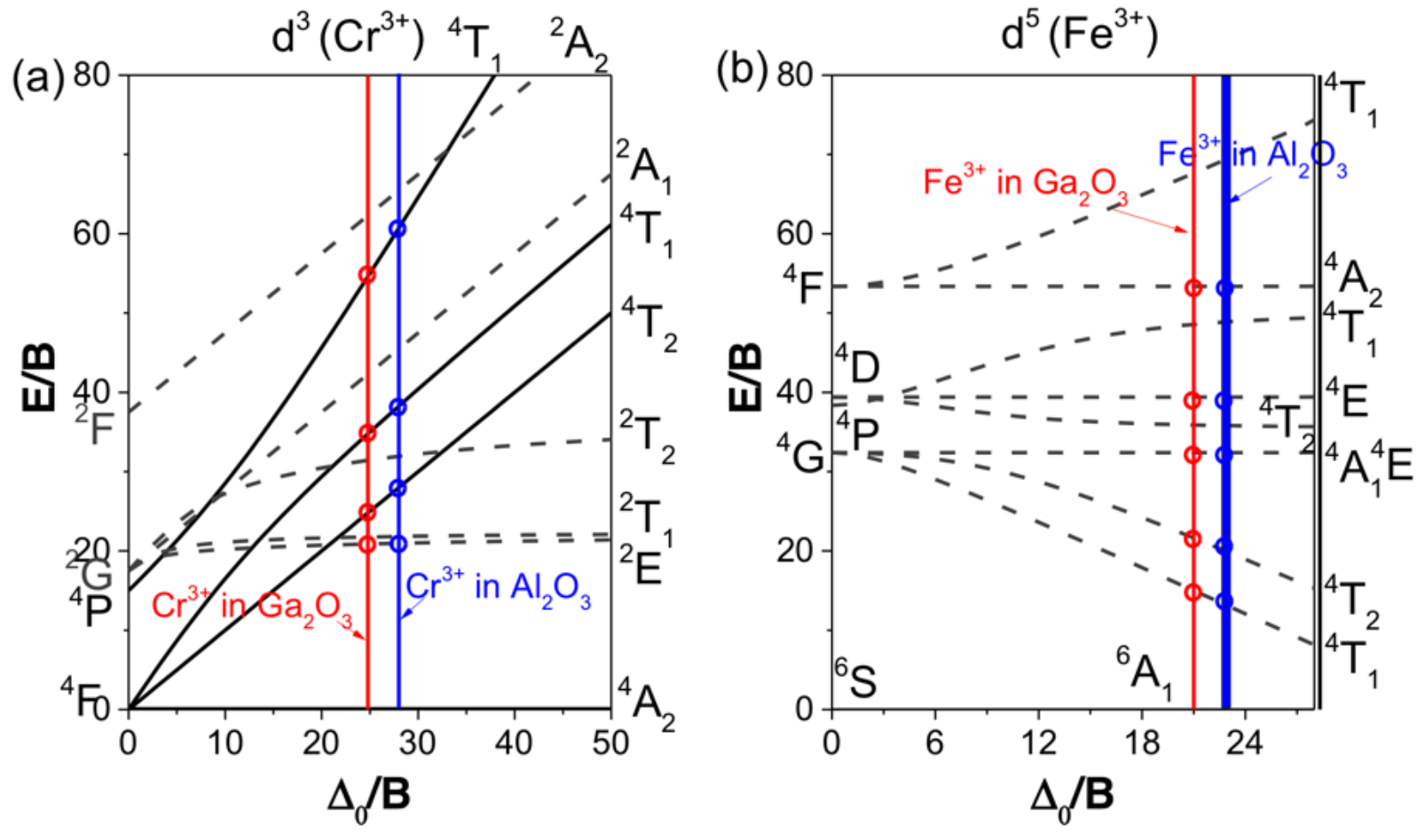

Figure 3 - Tanabe-Sugano diagrams for octahedral site of (a) $d^3$ and (b) $d^5$ electron configuration. The x-axis is $\frac{\Delta_0}{B}$ where $\Delta_0$ and B denote the crystal splitting energy and Rach parameter related to electron repulsion, respectively, and the y-axis is the transition energy E normalized to B. Solid and dash lines mean spin-allowed and spin-forbidden transitions respectively.

Hence, the red emission we observe from Fe-doped $\beta$-$Ga_2O_3$ likely originate from $Cr^{3+}$ instead of from $Fe^{3+}$ itself. The NCT (010) Fe-doped crystal was analyzed by ICPMS. We measured 12 ppm [Fe] (~$1.1\times10^{18}$ $cm^{-3}$), which is in agreement with the reported ~$8\times10^{17}$ $cm^{-3}$ [Fe] by GDMS in samples from the same vendor[9]. Our ICPMS did detect 2 ppm [Cr] corresponding to $1.8\times10^{17}$ $cm^{-3}$. Note that 0.3 ppm [Cr] is reported in UID crystals[30] and 0.4-0.6 ppm in our Zr-doped crystals. Approximately, ~ 6× higher [Cr] is observed in Fe-doped crystal. The effective segregation coefficient of Cr is larger than 1.0 (reported to be 1.45-3.11[49]) such that Cr from $Ga_2O_3$ feedstock tends to spontaneously concentrate into the growing crystal. Cr (along with Si and Fe) would also be expected to leach from the Ir crucible (99.99% purity is common) into melts, and ultimately into growing crystals.

We propose an energy transfer process between β-$Ga_2O_3$ and Cr, probably sensitized by Fe in Fe-doped β-$Ga_2O_3$, namely, electron-hole pairs generated in $\beta$-$Ga_2O_3$ recombine by transferring energy to excite $Fe^{3+}$ and $Cr^{3+}$. Excited $Fe^{3+}$ returns to its ground state by transferring energy to excite $Cr^{3+}$. The red luminescence occurs from excited $Cr^{3+}$. The lines of evidence we use to reach this conclusion are as follows: 1) the red emission intensifies with $E_{ex} \approx E_g$ for PLE and shows 2-fold symmetry which coincides with E//a*, b and c axes for polarized excitation, 2) the red emission competes with β-$Ga_2O_3$ matrix emission (red emission fades and β-$Ga_2O_3$ emission emerges with $E_{ex}$ > 4.90 eV in E//c) (Fig.2d) 3) near-resonant energy levels exist predicted by Tanabe-Sugano diagram, and 4) the red emission is clearly observed by naked eyes with ~30 mW power in Fe-doped $\beta$-$Ga_2O_3$ crystals but not measurable in UID or other intentionally doped $\beta$-$Ga_2O_3$ crystals. Such difference cannot be explained by ~ 6× higher Cr in Fe-doped $Ga_2O_3$ compared to UID one (roughly 0.5 ppm of Fe content in UID[30]). The following phenomena also support our hypothesis. A resonant energy transfer has been proposed in Cr-doped $\beta$-$Ga_2O_3$ crystals and films, evidenced by 1) blue emission is more reduced than UV and green emissions in higher Cr-doped $\beta$-$Ga_2O_3$ crystal[39] and 2) the red luminescence peaks with $E_{ex} \approx E_g$ and is less efficient with $E_{ex}$ > 5.06 eV[50]. In in α-$Ga_2O_3$ powder, Cr luminescence is observed to be nonlinearly enhanced by the coexistence of Fe [47]. In term of non-radiative process, deep state recombination centers at Fe or Cr, or something else unknown might also exist. More work is needed to illustrate the details of the energy transfer process in Fe-doped $\beta$-$Ga_2O_3$. But, our findings clearly demonstrate that the red luminescence we observe originates from Cr centers, not Fe ones.

In conclusion, polarized transmittance observed reduced optical band gap and increased sub-bandgap absorption. A red emission consisting of two sharp peaks at 689 nm and 697 nm superimposing a broad peak at 710 nm was observed from all Fe-doped $\beta$-$Ga_2O_3$ samples under deep UV excitation while the typical $\beta$-$Ga_2O_3$ PL spectrum from 300 to 600 nm was quenched. The red luminescence probably originated from $Cr^{3+}$, instead of Fe and N, based on low temperature PL and Tanabe-Sugano analysis. The red emission in E//c was inefficient above 4.9 eV excitation and $\beta$-$Ga_2O_3$ emission occurred simultaneously. We proposed an energy transfer process exists between $Ga_2O_3$ and Cr, probably sensitized by Fe in Fe-doped $\beta$-$Ga_2O_3$ for red luminescence.

**Data Availability**

The data that support the findings of this study are available from the corresponding author upon reasonable request.

**Acknowledgements**

This material is based upon work supported by the Air Force Office of Scientific Research under award number FA9550-18-1-0507 (Program Manager: Dr. Ali Sayir). Any opinions, findings, conclusions, or recommendations expressed in this material are those of the author(s) and do not necessarily reflect the views of the United States Air Force. The authors thanked Mr. John Blevins at AFRL for supplying us with the Synoptics Fe-doped $Ga_2O_3$ substrates and Prof. Steve Blair at the University of Utah for providing polarized optical measurement facilities. This work is dedicated to the memory of the late Prof. Kelvin G. Lynn.